\documentclass[reprint,amsmath,amssymb,aps,prb,twocolumn,showpacs]{revtex4}

\makeatletter
\usepackage{graphicx}
\usepackage{dcolumn}
\usepackage{bm}
\usepackage{amsmath,graphicx,latexsym}
\usepackage{verbatim}   

\def\P{\mathbf{P}}
\def\D{\mathbf{D}}
\def\e{{\mathcal E}}
\def\E{\bm{\mathcal E}}
\def\F{{\mathcal F}}
\def\PTO{PbTiO$_3$}

\begin{document}

\title{Mapping the energy surface of \PTO\ in multidimensional
electric-displacement space}

\author{Jiawang Hong}
\email{hongjw10@physics.rutgers.edu}
\affiliation{ Department of Physics and Astronomy, Rutgers University,
 Piscataway, NJ 08854-8019, USA }

\author{David Vanderbilt}
\affiliation{ Department of Physics and Astronomy, Rutgers University,
 Piscataway, NJ 08854-8019, USA }

\date{\today}

\begin{abstract}

In recent years, methods have been developed that allow first-principles
electronic-structure calculations to be carried out under conditions
of fixed electric field.  For some purposes, however, it is more
convenient to work at fixed electric displacement field.  Initial
implementations of the fixed-displacement-field approach have been
limited to constraining the field along one spatial dimension only.
Here, we generalize this approach to treat the full three-dimensional
displacement field as a constraint, and compute the internal-energy
landscape as a function of this multidimensional displacement-field
vector.  Using \PTO\ as a prototypical system, we identify
stable or metastable tetragonal, orthorhombic and rhombohedral
structures as the displacement field evolves along [001], [110] and
[111] directions, respectively.  The energy minimum along [001] is
found to be deeper than that along [110] or [111], as expected
for a system having a tetragonal ground state. The barriers
connecting these minima are found to be quite small, consistent
with the current understanding that the large piezoelectric effects
in \PTO\ arise from the easy rotation of the polarization vector.

\end{abstract}

\pacs{77.80.-e,71.15.-m}

\maketitle

\section{Introduction}
\label{sec:intro}

Since their introduction almost a decade ago,\cite{fixedE,umari}
methods for carrying out first-principles electronic-structure
calculations under conditions of fixed electric field $\e$
have found wide application in the study of the dielectric,
piezoelectric, and ferroelectric behavior of
materials.\cite{Max-fixE,Antons,Veithen,Fu-prl}
More recently, variants of this approach, in which the electric
polarization\cite{fixedP} or the electric
displacement field \cite{Max-NP} is taken as the fundamental
variable instead, have been introduced.  Fixing the displacement field $D$
has the intuitive interpretation of
imposing open-circuit electrical boundary conditions, which
is often especially useful for studying layered geometries
such as metal-oxide interfaces\cite{Max-prl,Max-prb-R} and
ferroelectric capacitors\cite{Max-PRB} and
superlattices.\cite{Xifan-prl,Xifan-prb}

There are several reasons why the fixed-$D$ approach is
advantageous for such applications.  In a superlattice
structure, the local polarization $P$ and electric field
$\e$ can vary from one layer to another, so their
overall spatial average is not a fundamental quantity.
In contrast, $D$ is constant throughout the system, since free
charge is assumed to be absent.  (Here, $\e$, $P$ and $D$
refer to the field components in the stacking direction.) Therefore,
choosing $D$ as the fundamental electrical variable is especially
practical because it makes it possible to decompose the equation
of state of a layered structure into the sum of contributions
from the individual building blocks,\cite{Max-PRB}
with each of these contributions depending only on the
local environment.  Moreover, long- and short-range electrostatic
interactions are separated efficiently.  Such an approach
facilitates the design of superlattice structures of desired
functionality by appropriate choice of the stacking
sequence.\cite{Xifan-prl,Xifan-prb}

Even for bulk crystals, it turns out that working at fixed
$\D$ is more convenient than working at fixed
$\E$ when considering materials that have
ferroelectric instabilities.  (We now refer to three-dimensional
field vectors.)  The reason is that, in the region of the energy
landscape near the unstable paraelectric configuration, the
system is unstable to transformation into one of the ferroelectric
domain states when working at fixed $\E$.  The unstable
paraelectric region of the $E(\P)$ energy landscape is thus
inaccessible using this approach.  When working at fixed $\D$,
on the other hand, the internal energy $U(\D)$ is typically found
to be a single-valued function of $\D$,\cite{explan-improper}
thus allowing access to the entire electric equation of
state.\cite{Max-NM} Moreover, the second derivative of $U$ with
respect to $\D$ is directly related to the inverse capacitance of
the material.\cite{Max-NP}  When this quantity goes negative, it
is a signature of the appearance of a ferroelectric instability,
and indeed the magnitude of its negative value has been shown to
be an insightful indicator of the strength of the ferroelectric
instability.\cite{Max-NP,Max-NM} For example, it can play an
important role in determining the critical thickness for
ferroelectricity in capacitor nanostructures,\cite{Max-PRB} and
its dependence on material structure and composition can be helpful
in understanding the origin of the ferroelectric instability.

Up to now, applications of the fixed-displacement-field
approach have been carried out using the private LAUTREC
code package,\cite{lautrec} in which the fixed-$D$ constraint
can be applied in only one Cartesian direction.  In the present
work, we have implemented the multidimensional fixed-$\D$ method
in the context of the open-source ABINIT code package\cite{Gonze}
and demonstrates that it can be used to calculate the energy
surface throughout the region of instability
associated with the ferroelectric behavior, using
\PTO\ as our prototypical material.  By mapping
the internal-energy landscape in the full three-dimensional
$\D$ space, we can easily compare the energies of the competing
ferroelectric states, trace the paths connecting these
states, and compute the energy barriers along these paths.
This approach therefore gives us a powerful tool for
characterizing the ferroelectric behavior of a given material
in all of its three-dimensional complexity in $\D$ (or $\E$
or $\P$) space.

The paper is organized as follows.  In Sec.~\ref{sec:method},
we briefly review the formalism for the fixed-$\D$ calculations
and describe our implementation. In Sec.~\ref{sec:results},
we first test the implementation and compare with a previous
calculation on \PTO.  We also explore the internal energy
landscape of \PTO\ in multidimensional $\D$ space, and 
map out the relationships between the various field variables
and their corresponding energy functionals.  Finally,
Sec.~\ref{sec:summ} contains a summary and conclusions.

\section{Formalism and methodology}
\label{sec:method}

We begin by briefly reviewing the fixed-$\E$ and fixed-$\D$
formalisms. For the former,\cite{fixedE,umari} the natural energy
functional is the electric enthalpy
\begin{equation}
\mathcal{F}(\bm{\mathcal{E}},v) = E_{\rm KS}(v) - \Omega\,
\bm{\mathcal{E}} \cdot \mathbf{P}(v),
\label{eq:enthalpy}
\end{equation}
while the fixed-$\D$ method\cite{Max-NP} is naturally formulated
in terms of the internal energy
\begin{equation}
U(\mathbf{D},v) = E_{\rm KS}(v) +
\frac{\Omega}{8\pi} [ \mathbf{D} - 4\pi \mathbf{P}(v) ]^2 \,.
\label{eq:functional}
\end{equation}
In these equations $\Omega$ is the cell volume, $v$ denotes
the internal (ionic and electronic) coordinates, and $E_{\rm KS}$
is the ordinary zero-field Kohn-Sham energy functional.

To implement the fixed-$\mathbf{D}$ method based on the above formalism,
we have modified the open-source ABINIT code package, in which
the fixed-$\bm{\mathcal{E}}$ calculation is already available.
\cite{fixedE,Gonze-E}
The key step is to update the
electric field $\bm{\mathcal{E}}$ after each SCF iteration
according to
\begin{equation}
\bm{\mathcal{E}}_{n+1}= \lambda (\mathbf{D}-4\pi \mathbf{P}_n) +
(1-\lambda) \bm{\mathcal{E}}_n ,
\label{damping}
\end{equation}
where $\mathbf{P}_n$ and $\bm{\mathcal{E}}_n$ are polarization
and electric field after the $n$'th SCF iteration and $\lambda$ is a
damping parameter used to control the convergence speed. The
iteration continues until the normal SCF convergence
criterion is reached, and in addition, $|\mathbf{D}-4\pi
\mathbf{P}_n-\bm{\mathcal{E}}_n|$ becomes less than a given
tolerance.

Our calculations were performed within density-functional theory
in the local-density approximation\cite{Perdew-Wang}
using norm-conserving pseudopotentials \cite{Norm-conserve} and a
plane-wave cutoff of 60 Ha.  The pseudopotentials were the same as
those of Ref~\onlinecite{Max-NP}.  A $6 \times 6 \times 6$
Monkhorst-Pack grid\cite{MP} was used to sample the Brillouin zone.
The atomic coordinates of the five-atom unit cell
were relaxed until all atomic force components
were smaller than 10$^{-5}$\,Ha/Bohr, and the cell size and shape was varied
until all stress components were below 10$^{-7}$\,Ha/Bohr$^3$.

\section{Results}
\label{sec:results}

\subsection{Testing consistency}
\label{sec:consistency}

In order to test our implementation and compare the results with the
previous calculation of Ref.~\onlinecite{Max-NP}, we have computed
the internal energy $U$ and the electric field $\e$ as a function
of $D$ for the case that $\D$ lies along the [001] axis.
We started the calculation from the relaxed cubic structure (lattice
constant of 3.88\,\AA) at $D=0$, and increased
$D$ in steps of 0.02\,a.u.  At each $D$,
the structure was fully relaxed with respect to both ionic positions
and lattice parameters.

\begin{figure}
  \begin{center}
    \includegraphics[width=2.8in]{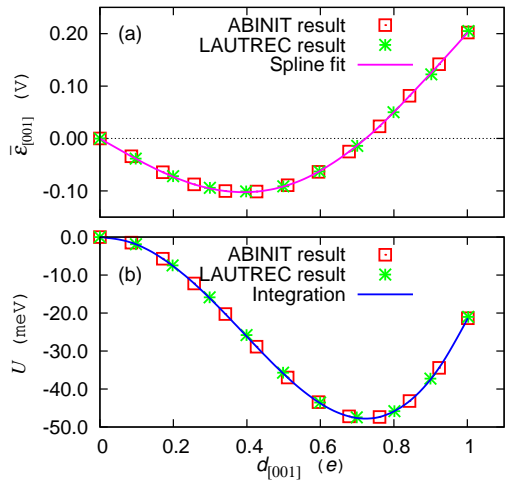}
  \end{center}
  \caption{(Color online) (a) Reduced electric field $\bar \varepsilon$,
  and (b) internal energy $U$, as a function of reduced electric
  displacement field $d$ along the [001] direction.  The LAUTREC
  results are reproduced from Ref.~\onlinecite{Max-NP}.  The
  solid curve in (a) is a cubic-spline fit to the ABINIT results;
  the solid curve in (b) is the numerical integral of the
  curve in (a).}
  \label{fig:p-u001}
\end{figure}

The results are plotted in Fig.~\ref{fig:p-u001}(a), except that
``reduced'' field variables are used.  That is, for each data
point we computed the reduced electric field
$\bar \varepsilon = c\e$ and the reduced displacement field
$d=a^2D/4\pi$, where $a$ and $c$ are the $x$-$y$ and $z$
lattice parameters respectively at the given value of $D$ along
[001].  In Fig.~\ref{fig:p-u001}(b), we plot the internal energy
$U$ vs.~reduced displacement field $d$.  The previous results of
Ref.~\onlinecite{Max-NP}, which were presented in terms of the same
reduced variables, are also included in the plots.
It is evident that the agreement between our Abinit calculation
and the previous Lautrec one is excellent.

An advantage of using reduced variables in Fig.~\ref{fig:p-u001}
is that $d$, $\bar \varepsilon$, and $U$ are related by the
exact relation
\begin{equation}
\frac{\partial U} {\partial d} = \bar \varepsilon
\label{eq:d-e}
\end{equation}
even when the relaxation of the lattice vectors with
$D$ is included.\cite{Max-NP}  (By contrast, the corresponding
relation between $D$, $\e$, and $U$ would require correction
terms arising from derivatives of lattice vectors and cell volume.)
To test whether our data is consistent with Eq.~(\ref{eq:d-e}),
we carried out a cubic spline fit of the $\bar \varepsilon$
vs.~$d$ results to produce the solid curve shown in
Fig.~\ref{fig:p-u001}(a), and then integrated it numerically
to obtain the solid curve shown in Fig.~\ref{fig:p-u001}b.  It can be
seen that the internal energies coming directly from the Abinit
calculations coincide quite precisely with those predicted by
the application of Eq.~(\ref{eq:d-e}).

These tests, which were repeated along other directions such as [110]
with equally good results, confirm the correctness and internal
consistency of our computational implementation.

\subsection{Internal energy in multidimensional $\D$ space}
\label{sec:multid}

\begin{figure}
  \begin{center}
    \includegraphics[width=2.8in]{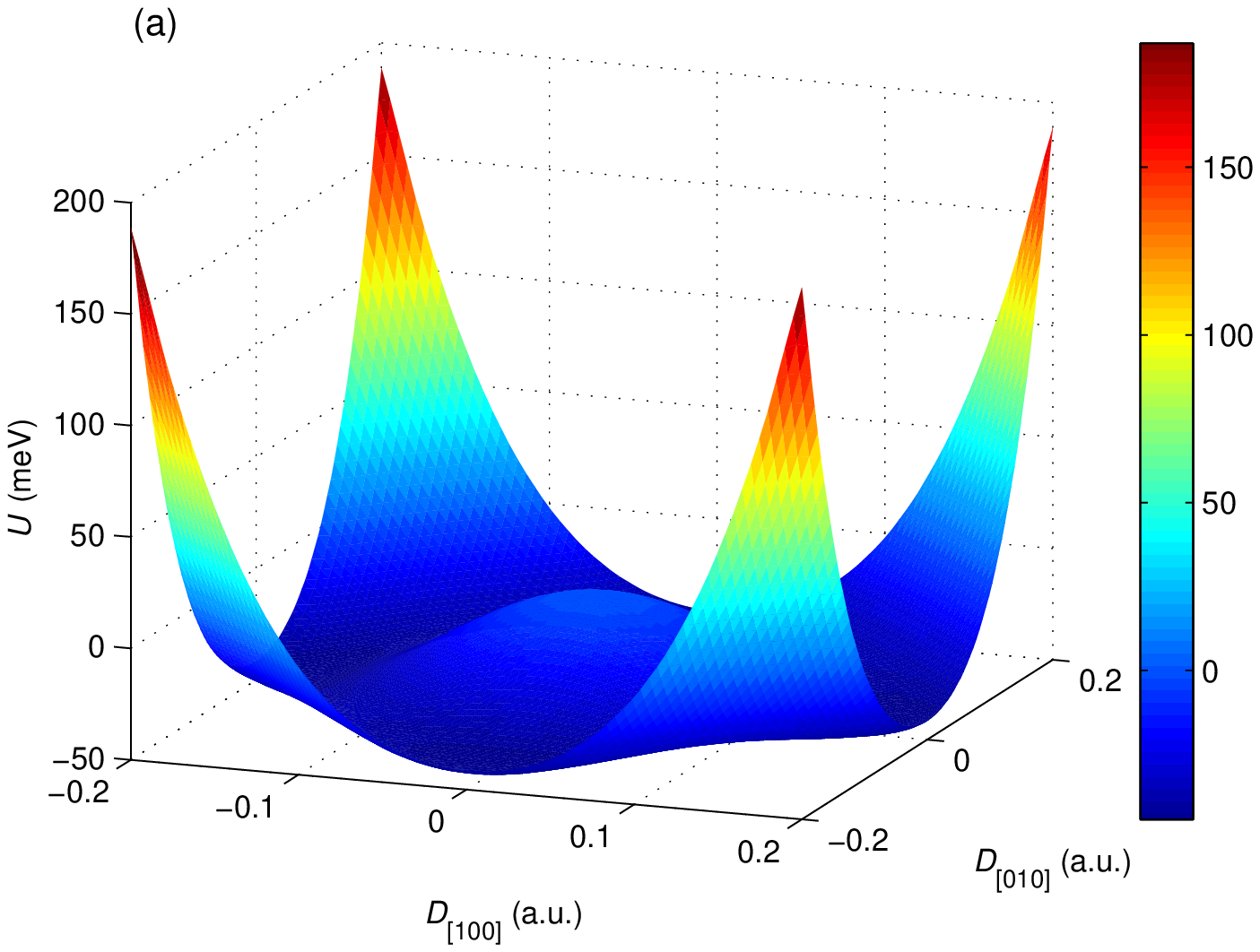}
    \includegraphics[width=2.8in]{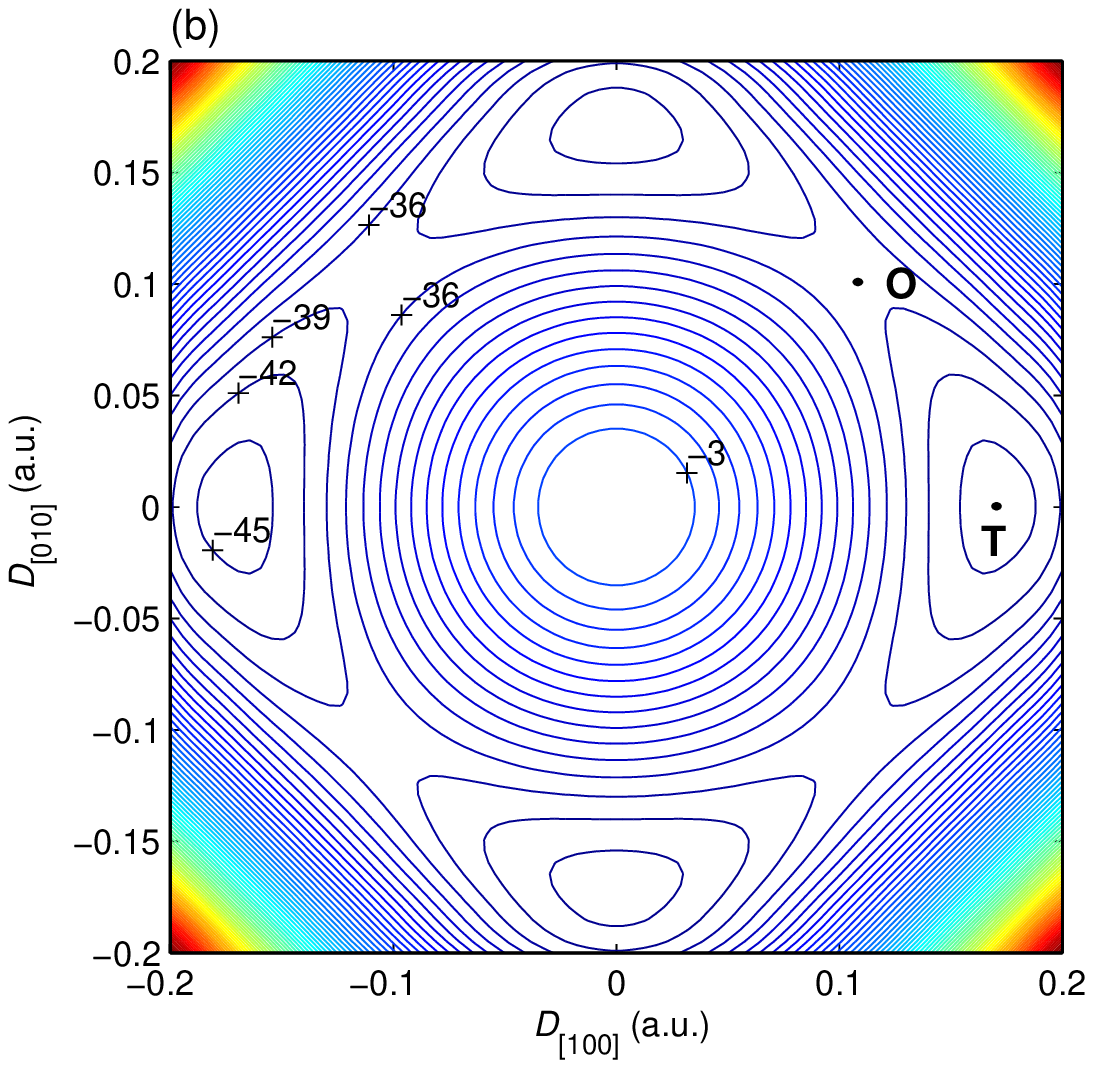}
  \end{center}
  \caption{(Color online) Internal energy surface $U(\D)$ of
  \PTO\ plotted for $\D$ lying in the plane spanned by
  the [100] and [010] directions. (a) Perspective plot.
  (b) Contour plot, with 3.0\,meV separating contour levels.
  The minimum at T and saddle point at O represent spontaneously
  polarized tetragonal and orthorhombic states respectively.}
  \label{fig:110_surface}
\end{figure}

We now explore the internal energy landscape of \PTO\ 
in three-dimensional $\D$ space. While our calculations can
be done for arbitrary $\D$, we restrict ourselves for
presentation purposes to two-dimensional planes in $\D$ space.
We begin with the case of $\mathbf{D}$ lying in the $x$-$y$ plane.
Recall that we already obtained relaxed solutions for a series
of $D_{[100]}$ values increasing in increments of 0.02\,a.u., up
to 0.2\,a.u., as described in the previous section.  For each of
these values of $D_{[100]}$, we use this solution as a starting
structure as we apply $D_{[010]}$ in steps
of 0.02\,a.u., up to 0.2\,a.u.,\ while keeping 
$D_{[100]}$ constant. In other words, each equilibrium state is
used as starting guess for its neighbor along $y$.
In this way $U(\D)$ is obtained on the specified grid in the quadrant
with $D_{[100]}>0$ and $D_{[010]}>0$, and then symmetry is used to
obtain the full internal-energy landscape.

The result is plotted in Fig.~\ref{fig:110_surface}.  Stationary
points in such a diagram correspond to states with $\E=0$.
The four minima in the $\pm$[001] and $\pm$[010] directions
correspond to four of the six tetragonal (T) ground
states, at which $P$ takes on its spontaneous value $P_{\rm s}$
and $D=4\pi P_{\rm s}$.  There are also four saddle points
along the [110] and related directions, corresponding to states
of orthorhombic (O) symmetry.  The fact that these are higher
in energy than the tetragonal minima just reflects the well-known
fact that \PTO\ has a tetragonal ground state at $T=0$.

\begin{figure}
  \begin{center}
   \includegraphics[width=2.8in]{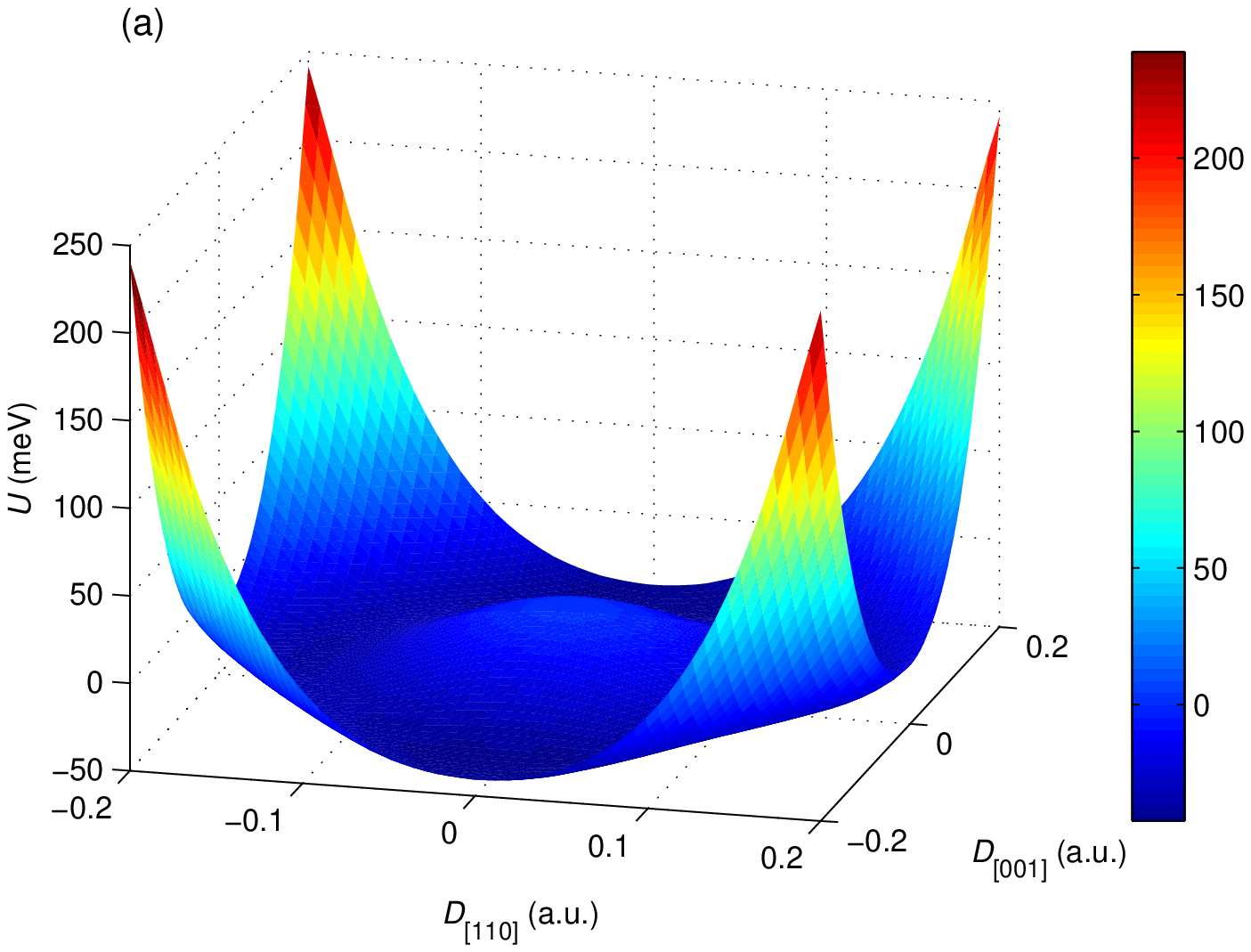}
   \includegraphics[width=2.8in]{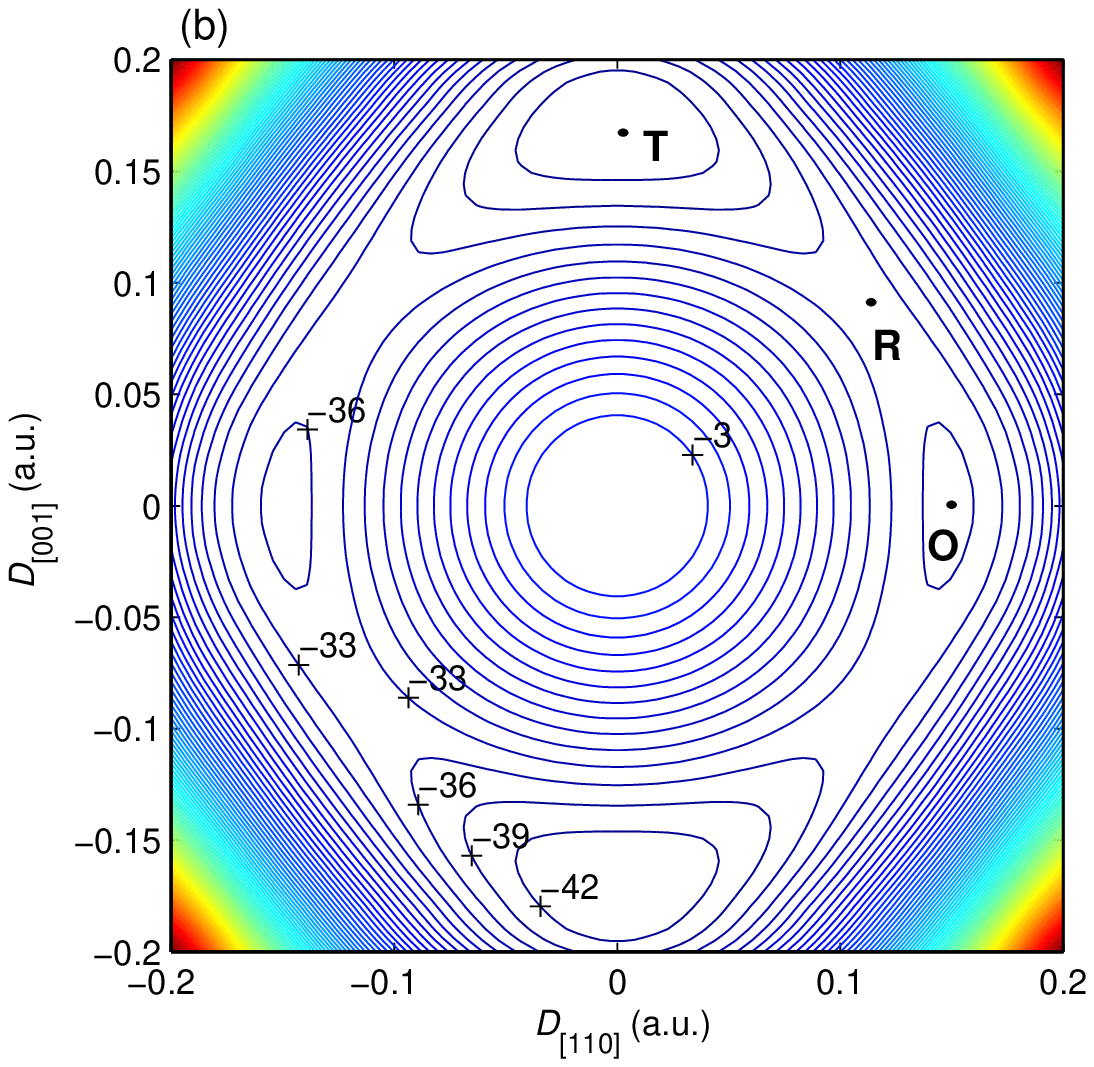}
  \end{center}
  \caption{(Color online) Internal energy surface $U(\D)$ of
  \PTO\ plotted for $\D$ lying in the plane spanned by
  the [110] and [001] directions. (a) Perspective plot.
  (b) Contour plot, with 3.0\,meV separating contour levels.
  The minima at T and O and the saddle point at R represent
  spontaneously polarized tetragonal, orthorhombic, and
  rhombohedral states, respectively.}
  \label{fig:110-001_surface}
\end{figure}

The figure also shows that the energy differences are quite small
along the valley connecting states T $\rightarrow$ O $\rightarrow$ T,
relative to the large barrier that would have to be overcome
if one were to pass through the origin of the figure.  This is
consistent with the work of Cohen and collaborators\cite{Fu,zhigang,Ganesh}
who pointed out how the easy rotation of the orientation of
the polarization could lead to large piezoelectric responses, even
if the energy landscape is relatively stiff with respect to changes in
the magnitude of the polarization.  Regarding the piezoelectric
response of \PTO\ starting from its tetragonal ground state, the
relative flatness of the energy landscape near state T in
Fig.~\ref{fig:110_surface} along that path that would lead from
T $\rightarrow$ O $\rightarrow$ T is heuristically consistent with
a relatively large observed value of the $e_{15}$ piezoelectric
constant in this material.\cite{zhigang}

Fig.~\ref{fig:110-001_surface} shows corresponding plots for the
plane spanning the [110] and [001] directions in three-dimensional
$\D$ space.  The minima at the top and bottom of panel (b) are
tetragonal states equivalent to those in Fig.~\ref{fig:110_surface}.
The apparent local minima at left and right in panel (b) are actually
saddle points in three-dimensional $\D$ space, and correspond to
the orthorhombic states already discussed in connection
with Fig.~\ref{fig:110_surface}.  We now also see four equivalent
saddle points corresponding to a rhombohedral (R) state with the
polarization along [111] or related directions; these are points
of double instability, in the sense that the Hessian of $U(\D)$
has two negative eigenvalues.
Once again, it is evident that there is a polarization rotation
path via T $\rightarrow$ R $\rightarrow$ O that is relatively
low in energy compared to a direct polarization reversal passing
through the paraelectric maximum at the origin.

\begin{figure}
  \begin{center}
   \includegraphics[width=2.8in]{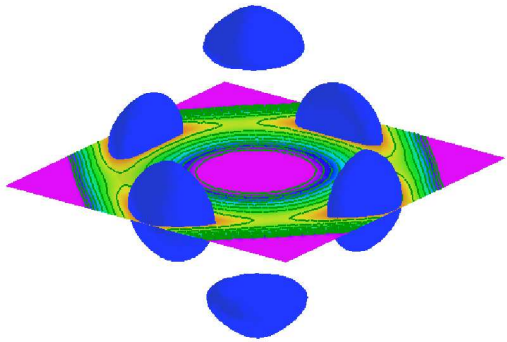}
  \end{center}
  \caption{(Color online) Dark blue: isosurfaces of $U(\D)$ for
  \PTO\ plotted at $U=-44$\,meV.  Contoured plane: alternative
  view of data of Fig.~\ref{fig:110_surface}(b).}
  \label{fig:3d}
\end{figure}

The properties of \PTO\ at the tetragonal, orthorhombic and
rhombohedral phases are summarized in Table~\ref{tab:TOR}.
The tetragonal phase has the lowest internal energy, followed
by the orthorhombic and then the rhombohedral phase, consistent with
previous studies and with the well-known fact that
the ground state of \PTO\ is tetragonal. It can also be seen
that the displacement field $D_{\rm min}$ minimizing $U$ and the
corresponding spontaneous polarization $P_{\rm s}=D_{\rm min}/4\pi$
decrease when going from T to O to R, as might have been guessed from
the fact that T is the deepest minimum while R is the shallowest.

To demonstrate the full 3D capabilities of the method, we also compute
$U(\D)$ on a 3D mesh of $\D$ values in increments of $\Delta
D_{[100]}=\Delta D_{[010]}=\Delta D_{[001]}=0.04$\,a.u.\ and use
this data to plot the internal-energy isosurface at $U=-44$\,meV
shown in Fig.~\ref{fig:3d}. The six isosurfaces surround
the six equivalent internal-energy minima (at $U=-47.3$\,meV)
located along the [001], [010] and [001] axes in $\D$ space,
providing a clear visualization of the six possible
spontaneously-polarized domain states in tetragonal \PTO.
The contour plane at $D_{[001]}=0$ shows similar information as
in Fig.~\ref{fig:110_surface}(b). These results demonstrate the
robustness of our implementation even in a fully multidimensional
context.

\subsection{Relations between field variables}
\label{sec:fields}

\begin{table}
\caption{Properties of \PTO\ in tetragonal (T), orthorhombic (O),
and rhombohedral (R) phases. The cubic phase is chosen to define the
zero of the internal energy $U$.  $D_{\rm min}$ is the displacement
field at which $U$ is a minimum, and $P_{\rm s}$ is the corresponding
spontaneous polarization.  The lattice vectors are also given.}
\label{tab:TOR}
\begin{ruledtabular}
\begin{tabular}{ccccc}
 & $U$ & $D_{\rm min}$ & $P_{\rm s}$ & $a$, $b$, $c$  \\
 & (meV/cell) & (a.u.)  & (C/m$^2$) & (\AA)   \\
\hline
T    & $-$47.78  & 0.17 & 0.78  &  $a=$3.85, $c=$4.03  \\
O    & $-$39.80 & 0.15 & 0.68  &  $a=b=$3.92, $c=$3.86 \\
R    & $-$37.23 &0.14  &0.65   &  $a$=3.90, $\alpha$=89.63$^\circ$  \\
\end{tabular}
\end{ruledtabular}
\end{table}

The electric equation of state of a given crystalline insulator
is given by specifying the relation between any two of the
three field variables $\P$, $\D$, and $\E$, as for example by the
functions $\E(\D)$, $\D(\P)$, or $\P(\E)$.  It is straightforward
to convert between these using $\D=\E+4\pi\P$.  In our approach
we obtain $\E(\D)$ directly, and then generate $\D(\P)$ or $\P(\E)$
by numerical manipulation.  In general the
electric equation of state is a vector function of a vector,
so to simplify our presentation we have calculated and plotted the
electric equations of state only for
cases in which all the field variables are constrained
to lie along either the [001], [110], or [111] axis.

\begin{figure*}
  \begin{center}
    \includegraphics[width=5.4in]{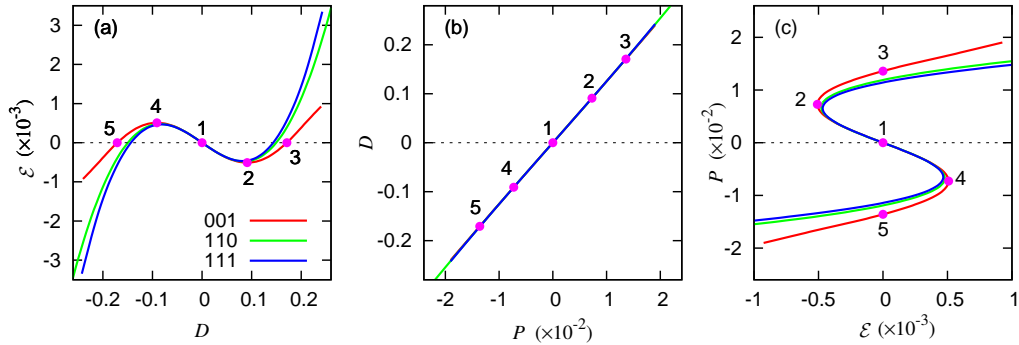}
  \end{center}
  \caption{(Color online) Electric equations of state of the
  form $\e(D)$ (a), $D(P)$ (b), and $P(\e)$ (c), plotted for
  fields constrained to lie along the [001], [110] or [111]
  directions.  All units are\,a.u. Numbered dots on the
  [001] curves indicate special states as described
  in the text.}
  \label{fig:variables}
\end{figure*}

The results are presented in Fig.~\ref{fig:variables}, with the
electric equations of state of the form $\e(D)$, $D(P)$, and
$P(\e)$ plotted in panels (a-c) respectively.  The different
curves correspond to plots along the [001], [110], or [111] axis.
We have also marked several special states on the diagrams
for the case of the fields being along the [001] axis.
Proceeding from 1 $\rightarrow$ 2 $\rightarrow$ 3 (or equivalently
by symmetry from 1 $\rightarrow$ 4 $\rightarrow$ 5), we
pass from the cubic paraelectric State 1 to the spontaneously
polarized tetragonal ferroelectric State 3.
As the displacement field $D$ increases in panel (b), the 
polarization $P$ increases nearly linearly, while the electric field
${\mathcal{E}}$ in panel (a) at first decreases to its
minimum value (which defines State 2) and then increases and passes through
zero at State 3.  The plot of $P(\e)$ in panel (c) takes
the form of a hysteresis curve, but the portions of this curve
in the region 4 $\rightarrow$ 1 $\rightarrow$ 2 are unstable and
therefore inaccessible under fixed-$\e$ boundary conditions.
This is a region in which the dielectric permittivity
$\chi=\partial P /\partial \mathcal{E}$ is negative.  The ability
of the fixed-$D$ method to explore this region of instability,
which cannot be done using the fixed-$\e$ method, is one of the
important advantages of working at fixed $D$.\cite{Max-NP}

Similar behaviors appear for $D$ along the [110] 
or [111] directions. For these directions, States 1-5 are not
marked, but are defined in the same way.
The region of instability from State 1 to 2
is almost the same along all three directions, as can be seen
in panels (a) and (c).  In all cases, $D$ is very nearly linear
in $P$ in panel (b).  The main difference comes in 
the $D$ and $P$ values at the spontaneously-polarized State 3,
which, as already discussed in connection with Table~\ref{tab:TOR},
increase as we go from the [111] (R) to the [110] (O) to the
[001] (T) directions.  Also, $\e$ increases much faster along
[110] and [111] than along [001] after
crossing the unstable State 2 in panel (a).

We emphasize that the hysteresis curves shown in
Fig.~\ref{fig:variables}(c) should not be compared directly with
experiment.  They correspond to the theoretical intrinsic
hysteretic behavior that would occur if the entire crystal
would switch coherently (i.e., maintaining full crystal
periodicity at all times) from $+P_{\rm s}$ to $-P_{\rm s}$
on a path crossing through $P=0$.  This is a highly unrealistic
picture of real ferroelectric switching, which usually proceeds
by the motion of a domain wall between domains of different
orientation of the polarization.~\cite{Beckman}
Our intrinsic coercive field of
2.5\,MV/cm for the [001] case, which can be obtained from $\e$
at State 2 in panel (a) or (c), is sure to be very much larger
than the experimental one, which is most likely determined by
pinning of domain walls to defects or by nucleation phenomena.

\subsection{Relations between energy functionals}
\label{sec:energies}

\begin{figure*}
  \includegraphics[width=5.4in]{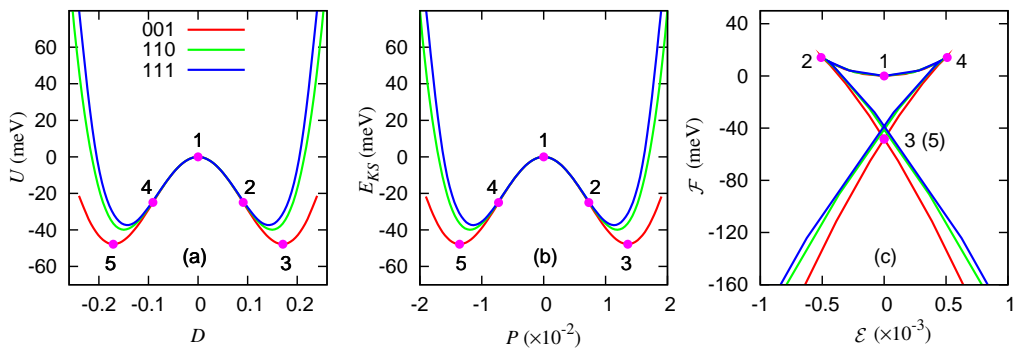}
  \caption{(Color online) Energy functionals vs.~their respective
  natural variables along [001], [110] and [111] directions:
  $U(D)$ (a), $E_{\rm KS}(P)$ (b), and $\F(\e)$ (c).  All field variables
  are in units of\,a.u. Dots labeled 1-5 indicate the same special
  states as in Fig.~\ref{fig:variables}.}
  \label{fig:energy}
\end{figure*}

It is also instructive to see how the energy functionals behave
as one traverses the trajectories shown in Fig.~\ref{fig:variables}.
Recall that the energy functionals that are naturally associated
with field variables $D$, $P$, and $\e$ are $U(D)$, $E_{\rm KS}(P)$,
and $\F(\e)$; these are plotted in panels (a-c) of
Fig.~\ref{fig:energy}, respectively.  Each plot again shows the
behavior for fields constrained along either [001], [110], or
[111], and the special States 1-5 are again marked for the [001]
case.

The $U(D)$ and $E_{\rm KS}(P)$ plots in panels (a-b) look remarkably
similar after a rescaling of the horizontal axis by a factor of
$4\pi$.  This is not surprising, since States 1, 3, and 5 have
$\e=0$, and are thus guaranteed to be extrema and
to appear in exactly the same place in both panels (after $4\pi$
rescaling).  The inflection points corresponding to States 2 and 4
are not located in quite the same place in both diagrams, but
it is difficult to tell this by eye.  Both panels clearly show
a qualitatively similar double-well potential.

The $\F(\e)$ curve in Fig.~\ref{fig:energy}(c) looks complex, but its
consistency with Fig.~\ref{fig:variables}(c) can be checked through
the relation $P= -\partial\F/\partial\e$, which follows from
Eq.~(\ref{eq:enthalpy}). Going through the unstable region from State 1
to State 2 in Fig.~\ref{fig:energy}(c), $\e$ becomes negative while
$-\partial\F/\partial\e$ becomes positive, consistent with the
corresponding behaviors of $\e$ and $P$ in Fig.~\ref{fig:variables}(c).
Then in the metastable region from State 2 to State 3, and into the
stable region beyond State 3, $\e$ returns to zero and then goes
negative while $-\partial\F/\partial\e$ continues to grow more
positive, again consistent with Fig.~\ref{fig:variables}(c).
The fact that $\F$ diverges to $-\infty$ as $|\e|$ becomes
large may look strange, but it is the normal behavior of
$\F(\e)$.  For example, for a simple linear dielectric, $U(D)$
and $E_{\rm KS}(P)$ are simple upright parabolas, while $\F(\e)$ is an
inverted parabola,\cite{Max-NP} and the asymptotic behaviors at large
$|\e|$ are similar to those appearing in Fig.~\ref{fig:energy}.

\section{Summary and conclusions}
\label{sec:summ}

In summary, we have demonstrated the possibility of carrying
out first-principles density-functional calculations under
boundary conditions in which all three components of the electric
displacement field $\D$ are fixed.  We have implemented the method
in the open-source ABINIT software package.

Using \PTO\ as a prototypical system, we have explored
the internal-energy landscape as a function of the full
three-dimensional displacement-field vector.  We have identified
the minimum-energy tetragonal, orthorhombic and rhombohedral structures,
and confirmed that the computed properties agree with previous
first-principles studies.  Our results allow for easy visualization
of the low-energy paths for polarization rotation, known to
be associated with large piezoelectric responses in this class
of compounds.  We have also presented the electric equations of
state relating $\e$, $D$ and $P$, as well as the corresponding energy
functionals, along symmetry lines in $\D$ space.

We hope that this fixed-$\D$ approach may be useful for exploring
the internal energy landscape of more complicated ferroelectric and
dielectric materials, in bulk or superlattice form, as well as for
studying domain-wall properties, piezoelectric and flexoelectric
responses, field-driven phase transitions, and other phenomena in
this important class of materials.

\acknowledgments

This work was supported by ONR Grant N00014-05-1-0054.
We thank P. Ganesh and E. Murray for the help with code
implementation. Computations were done at the Center for
Piezoelectrics by Design.


\begin{thebibliography}{99}
\bibitem{explan-improper} In our experience this is always
true for proper ferroelectrics, but not necessarily for
improper ones.
\bibitem{lautrec} LAUTREC is a parallel ``in-house''
electronic-structure code developed primarily by M.~Stengel.
\end{thebibliography}
\end{document}